\documentclass[
 prx,
 amsmath,
 amssymb,twocolumn,10pt,superscriptaddress
]{revtex4-2}

\usepackage{graphicx}% Include figure files
\usepackage{dcolumn}% Align table columns on decimal point
\usepackage{bm}% bold math
\usepackage{amsmath}
\usepackage{xcolor}

\DeclareUnicodeCharacter{2009}{ }

\begin{document}
	
	\title[]{Ultrafast Dynamics of Gallium Vacancy Charge States in $\beta$-Ga$_2$O$_3$}
	
	\author{Arjan Singh}
	\email{as2995@cornell.edu}
	
	\author{Okan Koksal}
	\affiliation{School of Electrical and Computer Engineering, Cornell University, Ithaca, New York 14853, USA.}
	
	\author{Nicholas Tanen}
	\affiliation{Department of Materials Science and Engineering, Cornell University, Ithaca, New York 14853, USA.}
	
	\author{Jonathan McCandless}
	\affiliation{School of Electrical and Computer Engineering, Cornell University, Ithaca, New York 14853, USA.}
	
	\author{Debdeep Jena}
	
	\author{Huili (Grace) Xing}
	\affiliation{School of Electrical and Computer Engineering, Cornell University, Ithaca, New York 14853, USA.}
	\affiliation{Department of Materials Science and Engineering, Cornell University, Ithaca, New York 14853, USA.}
	
	\author{Hartwin Peelaers}
	\affiliation{Department of Physics and Astronomy, University of Kansas, Lawrence, Kansas 66045, USA.}
	
	\author{Farhan Rana}
	\affiliation{School of Electrical and Computer Engineering, Cornell University, Ithaca, New York 14853, USA.}
	
	\date{\today}

	\begin{abstract}
	  Point defects in crystalline materials often occur in multiple charge states. Although many experimental methods to study and explore point defects are available, techniques to explore the non-equilibrium dynamics of the charge states of these defects at ultrafast (sub-nanosecond) time scales have not been discussed before. We present results from ultrafast optical-pump supercontinuum-probe spectroscopy measurements on $\beta$-Ga$_2$O$_3$. The study of point defects in $\beta$-Ga$_2$O$_3$ is essential for its establishment as a material platform for high-power electronics and deep-UV optoelectronics. Use of a supercontinuum probe allows us to obtain the time-resolved absorption spectra of material defects under non-equilibrium conditions with picosecond time resolution. The probe absorption spectra shows defect absorption peaks at two energies, $\sim$2.2 eV and $\sim$1.63 eV, within the 1.3-2.5 eV probe energy bandwidth. The strength of the absorption associated with each peak is time-dependent and the spectral weight shifts from the lower energy peak to the higher energy peak with pump-probe delay. Further, maximum defect absorption is seen for probe polarized along the crystal c-axis. The time-dependent probe absorption spectra and the observed dynamics for all probe wavelengths at all pump-probe delays can be fit with a set of rate equations for a single multi-level defect. Based on first-principles calculations within hybrid density functional theory we attribute the observed absorption features to optical transitions from the valence band to different charge states of Gallium vacancies. Our results demonstrate that broadband ultrafast supercontinuum spectroscopy can be a useful tool to explore charge states of defects and defect dynamics in semiconductors.
	\end{abstract}
	
	 % PACS, the Physics and Astronomy Classification Scheme.
	
	
	\maketitle
	
	% ----------	
	% motivation
    % ----------

\section{Introduction}        
    $\beta$-Ga$_2$O$_3$, an ultra-wide bandgap material, is a very promising candidate for high power electronic devices, solar-blind UV photodetectors, and sensors \cite{li_schottky_diode, hu_vertical_transistor, hu_uvdet, oshima_uvdet, ji_uvdet}. The availability of good quality large-area Ga$_2$O$_3$ substrates, the high breakdown electric field of the material, the ability to $n$-dope the material over a wide concentration range, the decent mobility of electrons, and the relatively long recombination times of photoexcited carriers in the material have all contributed to this promise. Most of these properties can be significantly impacted by material defects \cite{CVW_defects_review}. $\beta$-Ga$_2$O$_3$ can have many intrinsic and extrinsic point defects, including vacancies, interstitials, and impurities  \cite{fu_defects_review, tadjer_defects_review, mccluskey_defects_review}. The behavior of many of these point defects  is not well understood. Developing a better understanding of the properties of these defects is critical in realizing the material's promise.
    
    First-principles calculations have been instrumental in determining the formation energies, charge states, optical and thermodynamic transition energies, and the corresponding optical cross-sections of point defects in $\beta$-Ga$_2$O$_3$ \cite{dong_Vo_DFT, yao_Vo_DFT, varley_Vo_DFT, zacherle_DFT, varley_DFT, deak_DFT,Ho_PL_DFT, deak_DFT, Peel_DFT,Peelaers2016,ingebrigtsen_defect_proton_irradiation,varley_ga_complex}. Among the intrinsic point defects, Ga and O vacancies and interstitials have been theorized to have small formation energies. Ga vacancies, in particular, have the smallest formation energies in $n$-doped $\beta$-Ga$_2$O$_3$ grown under oxygen-rich conditions \cite{zacherle_DFT,deak_DFT,Peel_DFT,Ho_PL_DFT,ingebrigtsen_defect_proton_irradiation, varley_DFT}. In $n$-doped $\beta$-Ga$_2$O$_3$, Ga vacancy is a deep compensating acceptor and, depending on the Fermi level, it can be present in various charge states. Many different experimental techniques, including scanning probe and transmission electron microscopy \cite{johnson_ga_complex}, cathode- and photo-luminescence spectroscopy \cite{gao_depth_resolved_CL}, electron spin resonance spectroscopy \cite{kananen_esr_Vga,deek_esr}, and deep level transient spectroscopy \cite{Zhang_dlts} have been used to explore point defects in $\beta$-Ga$_2$O$_3$. However, none of these techniques have allowed simultaneous probing of different charge states of defects at ultrafast time scales. Since carrier capture by defects in $\beta$-Ga$_2$O$_3$ occurs on picosecond to nanosecond time scales \cite{koksal_opop}, it is important to be able to probe defect dynamics on these ultrafast time scales.

    In this paper, we present results from ultrafast optical-pump supercontinuum-probe spectroscopy of defects in $\beta$-Ga$_2$O$_3$, combined with first-principles calculations. Pump-probe spectroscopy is especially useful in exploring defects because, first, it allows for synchronized lock-in detection, which enables detection of fractional changes in light intensity as small as $10^{-6}$ \cite{wang_opop}. Such a degree of sensitivity is useful given the relatively small optical absorption due to defects. Second, pump-probe spectroscopy allows us to probe materials under non-equilibrium conditions. In $n$-doped materials, defect states are typically filled with electrons, disallowing optical transitions from the valence band (VB) to these defect states. In this work, we excite electrons out of the defect states using an optical pump pulse and probe the subsequent relaxation of the defect states, as they transition through different charge states toward their equilibrium state. We probe this defect relaxation using a broadband supercontinuum optical pulse, which is frequency filtered, to obtain the transient absorption by the defects at different energies. This yields the time-dependent absorption spectra of the defects with picosecond time resolution.

    \begin{figure}[htb]
		\includegraphics[width=1.0\columnwidth]{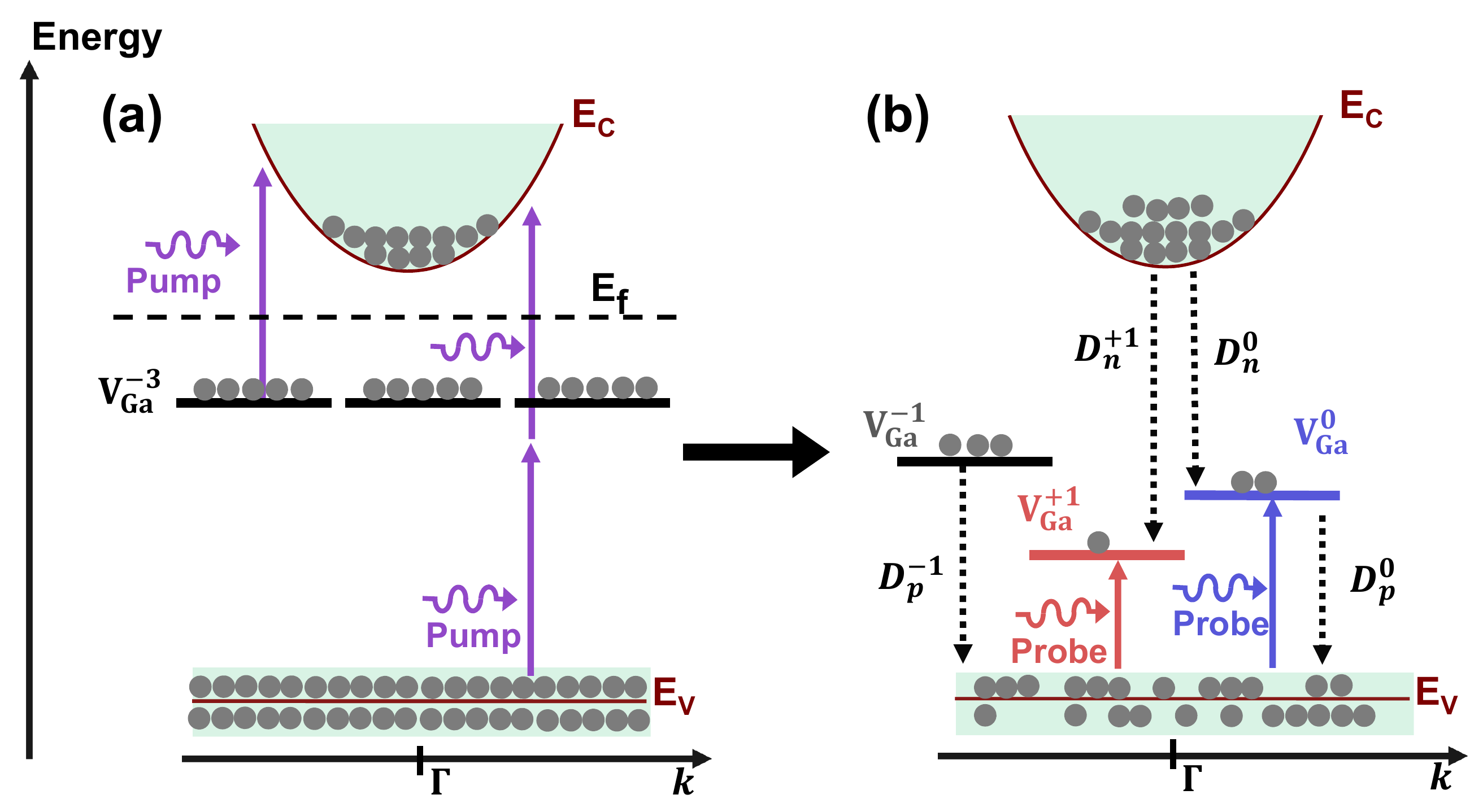}
		\caption{(a) The conduction and valence bands of $n$-doped $\beta$-Ga$_2$O$_3$ are depicted along with the optical excitation scheme. Also shown are the Ga vacancies in their equilibrium -3 charge state. (b) A snapshot of the non-equilibrium state some time after the optical excitation. The Ga vacancies are present in different charge states which allow optical transitions from the valence band and these transitions contribute to the transient absorption experienced by the probe pulses.}
		\label{fig:scheme}
	\end{figure}
        
    Our results show that transient optical absorption by defects in $\beta$-Ga$_2$O$_3$ depends sensitively on the polarization of the probe pulse. Absorption is maximum for the probe polarized along the crystal c-axis. The probe absorption spectra shows peaks at two energies, $\sim$2.2 eV and $\sim$1.63 eV, within the 1.3-2.5 eV probe energy bandwidth. Furthermore, we find the strength of the absorption associated with each peak to be time-dependent, with the spectral weight shifting from the lower energy peak to the higher energy peak with increasing pump-probe delay. The polarization dependence and energies of the defect absorption, and our ability to fit the time-dependent absorption spectra (and the observed temporal dynamics), for all probe wavelengths and all pump-probe delays with a set of rate equations for a single multi-level defect, allow us to attribute the observed absorption features to Ga(I) vacancies (or to defect complexes involving two Ga(I) vacancies and a Ga interstitial \cite{johnson_ga_complex,varley_ga_complex,ingebrigtsen_defect_proton_irradiation}). First-principles calculations show that the observed polarization dependence of the absorption features as well as the energies of the absorption peaks match well with optical transitions from the valence-band maximum to the +1 and 0 charge states of Ga(I) vacancies. Electrons are excited from the Ga(I) vacancies by the pump pulse thereby allowing optical transitions from the valence band to these vacancies. Fig.~\ref{fig:scheme} shows a depiction of our experimental scheme. While optical pump-probe spectroscopy has been used in the past to study relaxation dynamics involving defects, to the best of our knowledge, this is the first time that dynamics involving different charge states of a defect have been studied using ultrafast spectroscopy.
        
	% ------------	
	% introduction
	% ------------

\section{Experiments and results}    
    The samples studied in this work were obtained from the Tamura Corporation and consisted of bulk Sn-doped ($010$) $\beta$-Ga$_2$O$_3$ substrates, with an electron density, $n\approx5\times10^{18}$ $cm^{-3}$ and a thickness of $\sim$ 450 ${\mu}m$. The samples were grown by the EFG method \cite{kuramata_efg, aida_efg} in oxygen-rich conditions. A 405 nm ($\sim$ 3.1 eV) optical pump pulse was used to excite electrons into the conduction band through two-photon (non-linear) absorption, partially emptying the valence bands and the midgap defect states. The pump pulse was generated by frequency doubling an 810 nm ($\sim$ 1.53 eV), 66 fs, optical pulse generated by a $\sim$ 83 MHz repetition rate Ti:Sapphire laser. Part of the laser pulse was also used to generate a broadband supercontinuum pulse (bandwidth 1.3-2.5 eV) using a photonic crystal fiber (FemtoWhite 800). The supercontinuum pulse was frequency filtered to obtain the probe pulse with the desired center wavelength. The time-delayed (with respect to the pump pulse) probe pulse was used to record the transient optical absorption in the sample. The fluence values of the pump and probe pulses were kept fixed at $\sim$ 3.4 mJ/cm$^2$ and $\sim$ 1 $\mu$J/cm$^2$, respectively. 

	\begin{figure}[htb]
		\includegraphics[width=0.8\columnwidth]{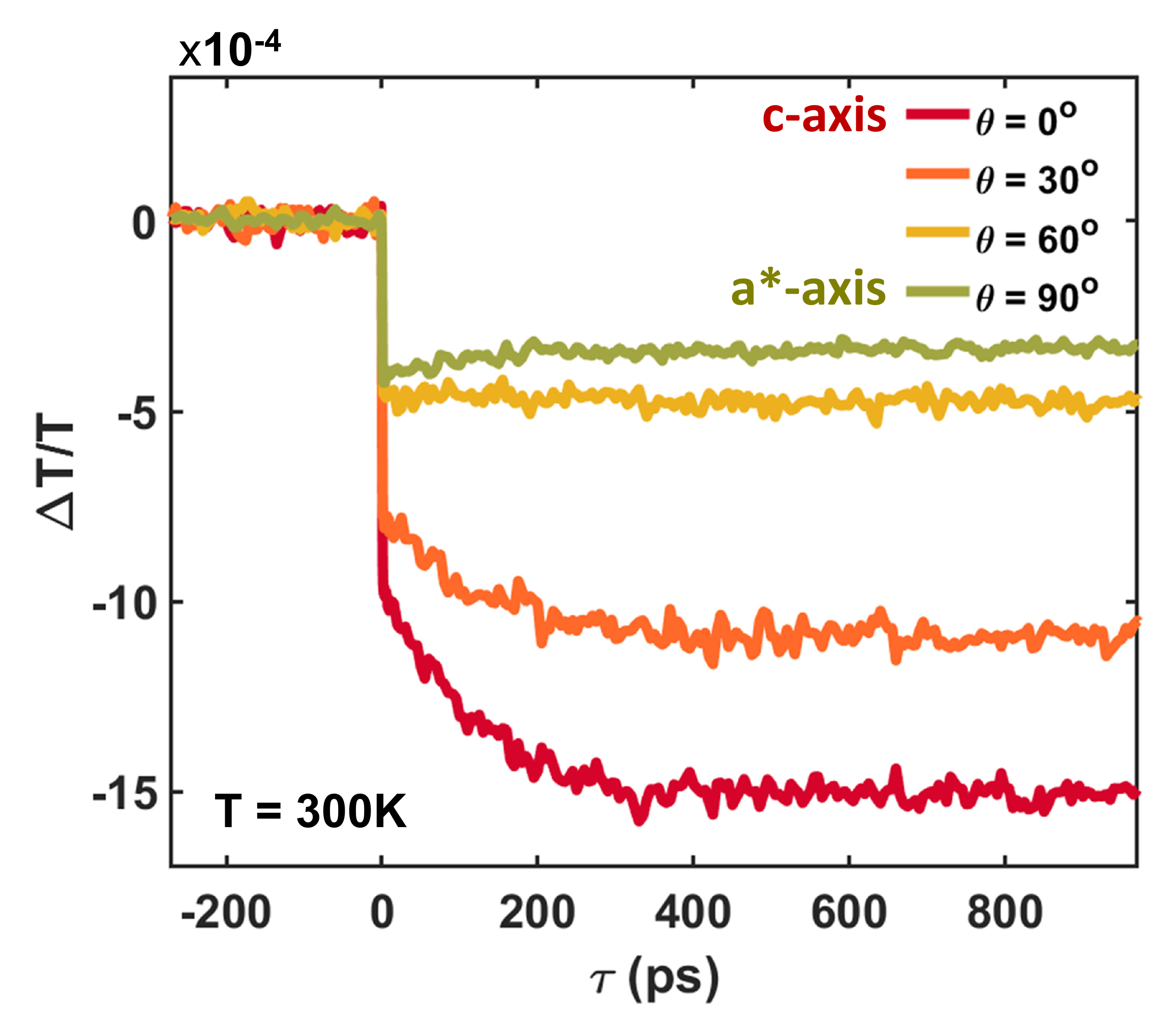}
		\caption{Differential probe transmission $\Delta T/T$ for n-doped ($010$) $\beta$-Ga$_2$O$_3$ is plotted as a function of probe delay. The probe wavelength is 800 nm. When the probe is polarized along the a*-axis (perpendicular to the b- and c-axis), the change in transmission is due to optical absorption from photoexcited free electrons (intra-conduction band absorption \cite{singh_fc}). As the polarization of the probe is changed away from the a*-axis toward the c-axis, additional absorption due to defects is observed that makes ${\Delta}T/T$ more negative in the first few hundred picoseconds.}
		\label{fig:pol_dep_800}
	\end{figure}

   	% -----------------------
	% polarization dependence
	% -----------------------
	
	Fig.~\ref{fig:pol_dep_800} shows the normalized differential transmission $\Delta T/T$, as a function of the pump-probe delay, of an 800 nm probe pulse polarized along different crystal axes for $(010)$ $\beta$-Ga$_2$O$_3$. The observed negative values of $\Delta T/T$ signify an increase in the optical absorption induced by the photoexcitation of electrons by the pump pulse. As seen in the figure, the measured $\Delta T/T$ is highly polarization dependent. Much larger peak values of $|\Delta T|/T$ are observed when the probe is polarized along the c-axis, and the peak values steadily decrease as the probe polarization is changed to be along the orthogonal a*-axis. Very notably, the shape of the $\Delta T/T$ transient is also polarization dependent suggesting that different loss mechanisms are contributing to the probe absorption when the probe is polarized along the c-axis and a*-axis. In recently reported work \cite{singh_fc}, we have examined the $\Delta T/T$ transient for probe polarization along the a*-axis in detail and shown that, for this polarization, the probe experiences optical loss only due to intra-conduction band transitions (a form of free-carrier absorption) characterized by a $1/\omega^{3}$ frequency dependence, where $\omega$ is the center frequency of the probe pulse. Optical absorption related to defects is not observed for probe polarized along the a*-axis. When the probe is polarized away from the a*-axis, we observe additional absorption (i.e. in addition to free-carrier absorption) that keeps increasing with the pump-probe delay for the first few hundred picoseconds. We attribute this additional absorption to optically active defect states. As can be seen in Fig.~\ref{fig:pol_dep_800}, defect absorption is maximum for probe polarized along the c-axis.
	% -----------------
	% defect absorption
	% -----------------
	
\begin{figure}[htb]
		\includegraphics[width=1\columnwidth]{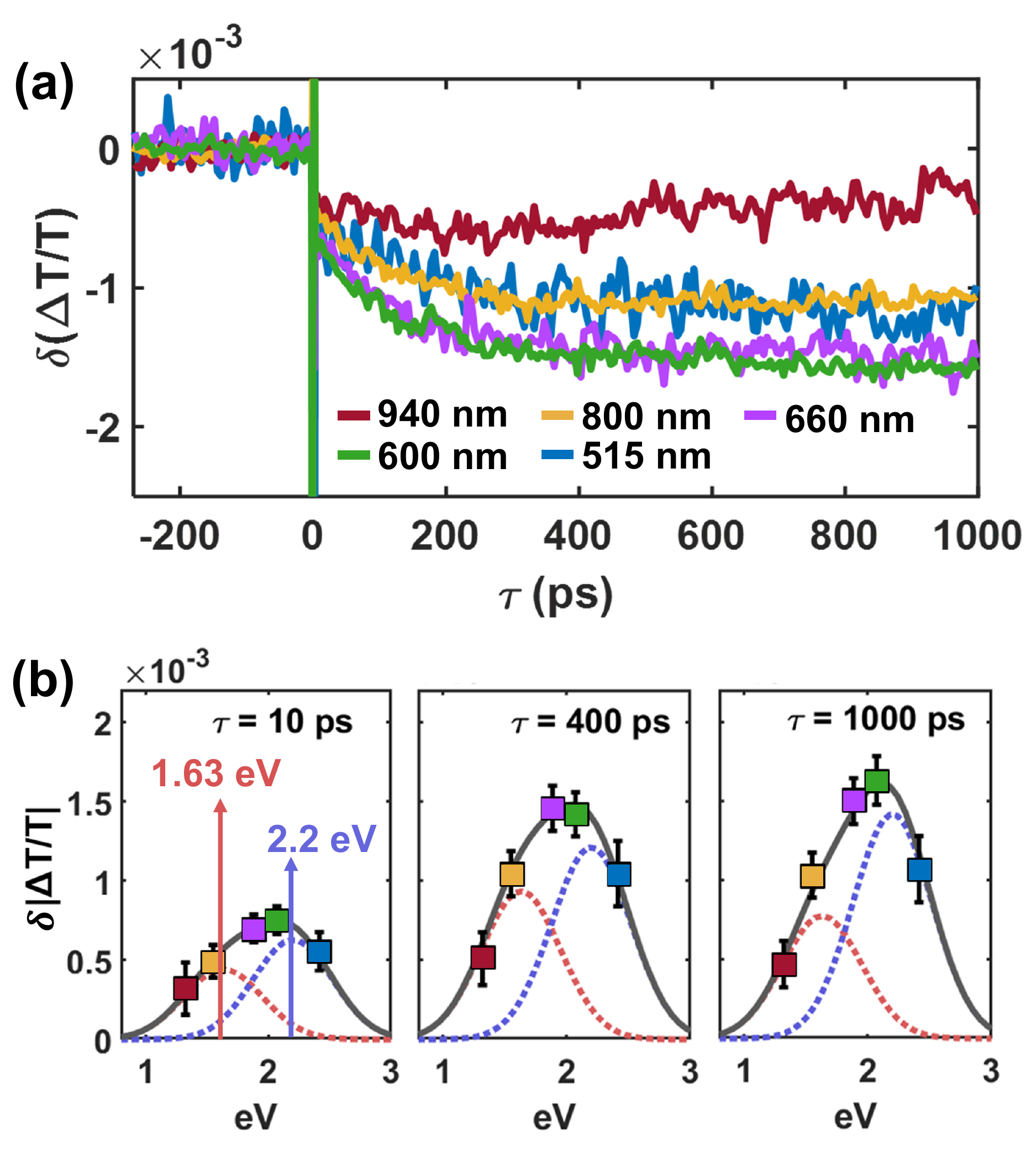}
		\caption{(a) The difference $\delta ({\Delta}T/T)$ between the ${\Delta}T/T$ transients measured for probe polarized along the c-axis and a*-axis is plotted as a function of the probe delay, for different probe wavelengths. (b) The defect absorption spectra (proportional to the magnitude of $\delta ({\Delta}T/T)$) are shown for different probe delays. The defect absorption spectra for all probe delays can be fitted using two Gaussian absorption coefficients centered at 1.63 eV ($\sim$ 761 nm) and 2.2 eV ($\sim$ 564 nm). The relative weights of these two Gaussian absorption coefficients change with time (but their widths stay constant).}
		\label{fig:defect_transmission}
	\end{figure}

	In order to better quantify the defect absorption, we measure $\Delta T/T$ for different probe wavelengths. Since free-carrier (intra-conduction band) absorption is expected to be polarization independent \cite{peelaers_ga2o3fc}, we subtract the measured ${\Delta}T/T$ along the a*-axis from that along the c-axis to obtain the defect absorption contribution to $\Delta T/T$. We refer to this modified normalized differential transmission as $\delta (\Delta T/T)$. $\delta (\Delta T/T)$ as a function of pump-probe delay, for various probe wavelengths is shown in Fig.~\ref{fig:defect_transmission}(a). Interestingly, the shapes of the  $\delta (\Delta T/T)$ transients are wavelength dependent (i.e. $\delta (\Delta T/T)$ transients for different wavelength are not just scaled versions of each other). The corresponding time-dependent defect absorption spectra for different probe delays are shown in Fig.\ref{fig:defect_transmission}(b). As can be seen in this Figure, the defect absorption spectra (proportional to the magnitude of $\delta ({\Delta}T/T)$) for all probe delays) can be fit using two Gaussian absorption coefficients, one centered at 1.63 eV ($\sim$ 761 nm) and the other at 2.2 eV ($\sim$ 564 nm). The relative weights of these two Gaussian absorption coefficients change with time (but the widths remain constant), resulting in the wavelength dependent temporal dynamics seen in Fig.~\ref{fig:defect_transmission}(a). Note that absorption outside the 1.3-2.5 eV bandwidth of our supercontinuum probe, which could be due to optical transitions from the valence band to the -1 and -2 charge states of Ga(I) vacancies (discussed below), is not detected in this work.  

	% --------------
	% Ga(I) Vaccancy
	% --------------

\section{Discussion}
The experimental observations allow us to conclude the following. First, the polarization dependent defect absorption in Fig.~\ref{fig:pol_dep_800} has been observed previously in both heavily and mildly $n$-doped $\beta$-Ga$_2$O$_3$ samples of different crystal orientations \cite{koksal_opop}, and is therefore unrelated to doping. Second, the relatively large strength of the absorption (proportional to the maximum magnitude of $\delta ({\Delta}T/T)$, which is of the order of 10$^{-3}$) signifies a fairly large concentration of the defects. Third, the increase in absorption in the first few hundred picoseconds after the pump pulse points to optical transitions from the valence band to the defect states being responsible for the defect-related optical absorption. Therefore, in the discussion that follows, {\em optical transition} will refer to the process in which an electron transitions from the valence band to a defect state after absorbing light. Fourth, the probe wavelength-dependent transients point to either multiple defects with different absorption spectra but the same polarization selection rule or to a single defect with multiple charge states.

\subsection{The nature of defect states: first-principles calculations}

The experimental data was analyzd with the help of first-principles calculations. We used density functional theory as implemented in the \textsc{VASP} code~\cite{Kresse1996}, using projector augmented wave potentials~\cite{Blochl1994} with an energy cutoff of 400 eV and a $2\times2\times2$ {\bf k}-point grid in a 120-atom 1$\times$3$\times$2 supercell (based on the conventional unit cell~\cite{Peelaers2015b}). To obtain accurate structural and electronic properties we used the HSE06 hybrid functional~\cite{Heyd2003,*Heyd2006} with a mixing parameter of 32\%. We used the defect formation energy formalism as outlined in Ref.~\onlinecite{Freysoldt2014}, with optical transition energies corrected by the scheme of Ref.~\onlinecite{Gake2020}. 

    Based on reported first-principles calculations we can exclude oxygen vacancies ($V_{\rm OI}, V_{\rm OII}, V_{\rm OIII}$), as these behave as deep donors, with the +2 to 0 thermodynamic transition occurring at Fermi energies larger than 2.5 eV (measured from the valence-band maximum)~\cite{varley_DFT,Peel_DFT,ingebrigtsen_defect_proton_irradiation}. Note that thermodynamic transition energies and optical absorption energies are not the same because the former include the effect of full lattice relaxation, which decreases the thermodynamic transition energy with respect to the optical absorption energy. Although the +1 charge state of an oxygen vacancy is not stable, optical transitions from +2 to +1 or from +1 to 0 charge states are possible but require larger photon energies than the ones observed in our experiments~\cite{varley_DFT, Yao2019a}. Ga vacancies, on the other hand, have characteristics that match all our experimental observations. A large concentration of Ga vacancies is expected to be present in our samples because of their very low formation energies in $n$-doped $\beta$-Ga$_2$O$_3$ \cite{ingebrigtsen_defect_proton_irradiation, deak_DFT, zacherle_DFT, varley_DFT}. In particular, the Ga(I) divacancy-interstitial complex ($V_{\rm Ga}^{ic}$ in the notation of Ingebrigtsen \textit{et al.}~\cite{ingebrigtsen_defect_proton_irradiation}) has the lowest formation energy of all intrinsic defects.

	\begin{figure}[htb]
	\includegraphics[width=1\columnwidth]{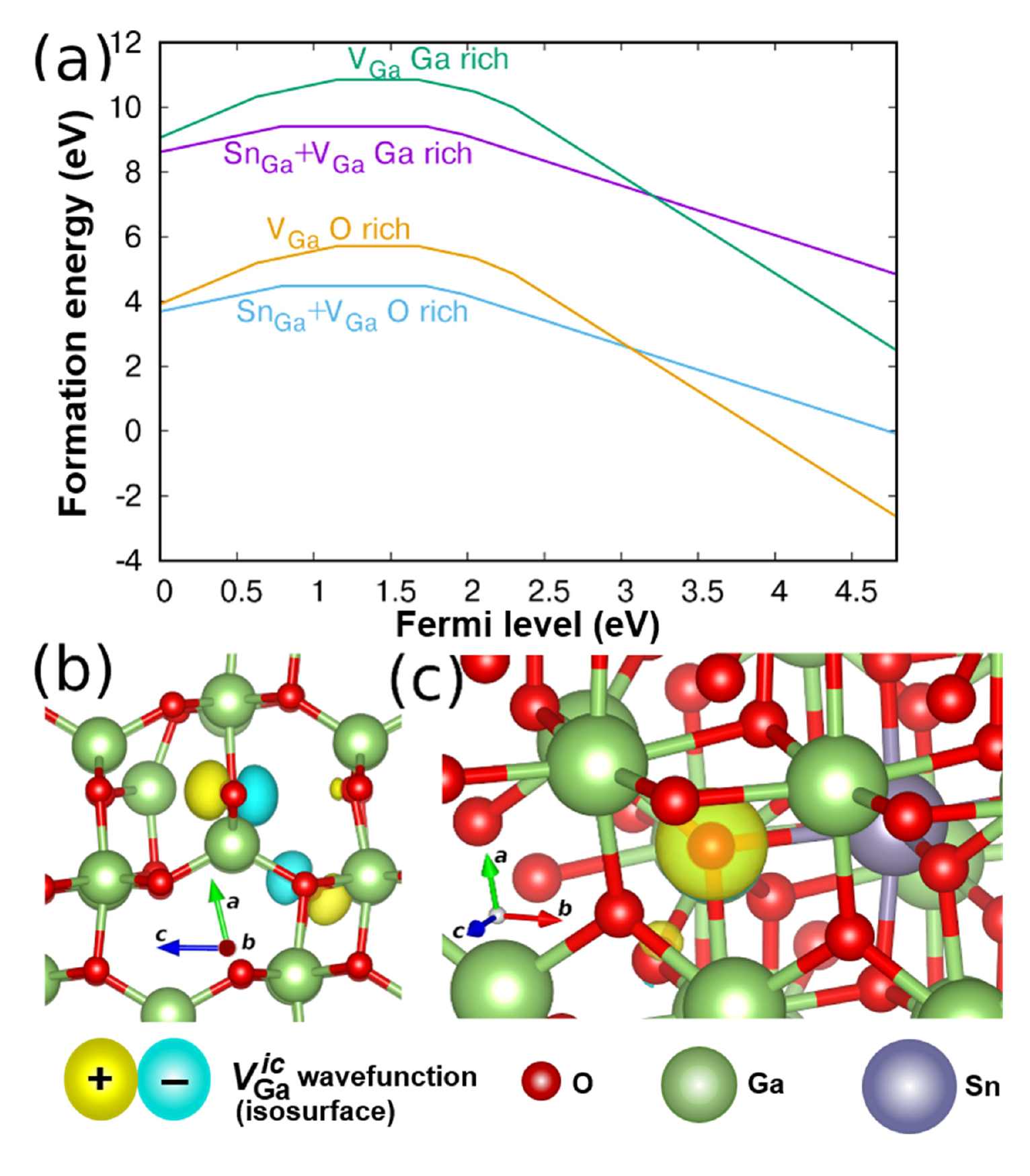}
	\caption{(a) Formation energies as function of Fermi level for $V_{\rm Ga}^{ic}$ and Sn-$V_{\rm Ga}$ complexes. The kinks in the curves indicate the thermodynamic transition levels. Both Ga-rich and O-rich conditions are shown. (b-c) Wavefunction isosurface at 10\% of the maximum value corresponding to the unoccupied defect state of the +1 charge state of the (b) $V_{\rm Ga}^{ic}$ complex and (c) the Sn-$V_{\rm Ga}$ complex.}
	\label{fig:comp}
    \end{figure}
        
	Ga vacancies can readily form complexes with other atoms and defects, such as hydrogen~\cite{varley_DFT}. We find that the absorption features seen in Fig.~\ref{fig:pol_dep_800} and Fig.~\ref{fig:defect_transmission} do not change upon annealing at 1100 K in 80\% O$_{2}$ ambient. We therefore exclude the possibility of hydrogenated Ga vacancies being responsible for these absorption features. In our Sn-doped samples, Ga vacancies can also form complexes with Sn dopants \cite{johnson_ga_complex}. The calculated formation energy diagram (Fig.~\ref{fig:comp}(a)) indicates that these Sn-$V_{\rm Ga}$ complexes can readily form and that they have thermodynamic transition levels at energies similar to those of Ga vacancies. Our first-principles calculations show that optical transitions from +1 to 0 and from 0 to -1 charge states of the $V_{Ga}^{ic}$ complex take place at 1.8 eV and 2.5 eV, respectively, both of which are very close to the experimentally observed absorption energies. The corresponding transitions for the complex with Sn are 1.7 eV  (+1 to 0) and 2.7 eV (0 to -1). To further distinguish between these two defects, we calculated the light polarization dependence expected for optical transitions from the valence band to the defect states. For these charge state transitions, the $V_{Ga}^{ic}$ complex shows a strong polarization dependency, with light polarized along the c-axis leading to the strongest absorption in agreement with our experiments. In contrast, the absorption for the Sn-$V_{\rm Ga}$ complex does not depend on the polarization with respect to the $a$, $b$, and $c$ crystal axes. The difference can be understood by looking at the wavefunction of the unoccupied defect state (+1 charge state): for the $V_{\rm Ga}^{ic}$ complex (Fig.~\ref{fig:comp}(b)) the wavefunction is mainly oriented along the c-axis, while for the Sn-$V_{\rm Ga}$ complex (Fig.~\ref{fig:comp}(c)) the wavefunction is not oriented along any of the crystal axes. This polarization dependence allows us to exclude the Sn-$V_{\rm Ga}$ complexes, and strengthens our conclusion that the measured absorption is due to the charge states of the Ga(I) vacancy.
	
	Our calculations show that optical transitions from -1 to -2 and from -2 to -3 charge states of the $V_{Ga}^{ic}$ complex are nearly polarization independent and do not show a preference for light polarization along the $c$-axis. We can therefore exclude these transitions playing a dominant role in causing probe absorption in our experiments. Finally, our calculations show that optical transitions from +2 to +1 charge states of the $V_{Ga}^{ic}$ complex do show a strong preference for light polarization along the $c$-axis but the calculated absorption energy of 1.4 eV for this transition is smaller than the measured energies. We therefore conclude that the probability that the pump pulse leaves the defect in charge state +2 is low. A possible explanation for this is as follows.
	
	Our earlier work~\cite{koksal_opop} showed that optical transitions caused by a $\sim$ 3 eV pump pulse from these defect states to the conduction band are due to a two-photon absorption process. This is consistent with our first-principles calculations for $V_{Ga}^{ic}$ to conduction band transitions, which give energies of 4.2, 4.6, and 5.2 eV for -1 to 0, 0 to +1, and +1 to +2 transitions, respectively. In these transitions, an electron is excited from the defect to the conduction band via a single-photon absorption process. Since these calculated energies are much larger than the $\sim$ 3.06 eV pump photon energy in our experiments here, our pump pulse is causing optical transitions from the defect charge states to the conduction band through a two-photon absorption process . In most semiconductors, if the energy required for a single-photon absorptive transition into the conduction band is $E_{o}$, then the strength of the corresponding two-photon transition can be approximately described by a universal function that peaks when the photon energy equals $\sim 0.71 E_{o}$ and rapidly approaches zero when the photon energy equals $\sim 0.5 E_{o}$~\cite{Boyd08}. It follows that the pump used in our experiments is much more likely to cause -1 to 0 and 0 to +1 charge state transitions than cause the +1 to +2 transition. Furthermore, as discussed below, a minimum model based on probe-induced optical transitions from +1 to 0 and from 0 to -1 charge states of the $V_{Ga}^{ic}$ complex can explain our data well. It is therefore safe to conclude that our pump pulse is not likely to put the $V_{Ga}^{ic}$ complex into the +2 charge state. 
		
	% -------
	% Fitting
	% -------

 \subsection{A rate equation model for the defect state dynamics}
    Next, we present coupled rate equations for modeling the non-equilibrium dynamics of the charge states of Ga vacancies and show that the computed wavelength-dependent and time-dependent $\Delta T/T$ transients, assuming that the defect optical absorption is due to the charge states of Ga vacancies, agree very well with our measurements. The probe frequency dependent ${\Delta}T/T$ can be written as,
	
	\begin{eqnarray}
	  	({\Delta}T/T) & = & e^{-\left[n(\tau)-n_{o}\right] {\sigma}_{fc}(\omega) L_i - n_{d} \sigma_d(\omega) f L_i} - 1 \nonumber \\
        & \approx & -\left[n(\tau)-n_{o}\right] {\sigma}_{fc}(\omega) L_i - n_{d} \sigma_d(\omega) f L_i
		\label{eq:deltaT/T}
	\end{eqnarray}
	
	where, $L_i$ is the pump-probe interaction length determined by the spatial overlap of the pump and probe beams in the sample ($\approx$ 260 $\mu$m in our measurement scheme), $n_o$ is the equilibrium electron density ($\sim5\times10^{18}$ $cm^{-3}$), $n(\tau)$ is the total electron density in the conduction band at time $\tau$, $\sigma_{fc}(\omega)$ is the absorption cross-section associated with free-carrier intra-band absorption \cite{singh_fc}. As shown recently by the authors \cite{singh_fc,peelaers_ga2o3fc}, $\sigma_{fc}(\omega)$ is proportional to $\omega^{-3}$, where $\omega$ is the frequency of the probe. $n_{d}$ is the defect density. $\sigma_d(\omega)$ is the defect absorption cross-section. $f$ equals 1 (or 0) for probe polarization along the c-axis (or a*-axis). $\sigma_d(\omega)$ can be written as, $\sigma_d(\omega)  =  \sum \sigma_{j}(\omega) P_{j}(\tau)$. $P_j(\tau)$ is the time-dependent probability of a Gallium vacancy being in the $j$ charge state. $\sigma_{+1}(\omega)$ and $\sigma_{0}(\omega)$ are the defect absorption cross-sections when the defect is in the $+1$ and $0$ charge states, and are assumed to be Gaussians centered at 1.63 eV and 2.2 eV, respectively. The widths of the Gaussians are chosen to fit the measured defect absorption spectra (see Fig.~\ref{fig:defect_transmission}), and the peak values of the Gaussians are used as fitting parameters. Note that, $\delta({\Delta}T/T) \approx  - n_{d} \sigma_d(\omega) L_i$. $P_j(\tau)$ are calculated using a defect-assisted carrier recombination rate equation model,	
	\begin{eqnarray}
		\frac{dn}{d\tau} & = & - \sum_{j} D^{j}_{n} \: n \: n_d \: P_{j}   \nonumber \\
		n_d \frac{dP_{j}}{d\tau} & = & - \left( D^{j}_n \: n \: n_d +   D^{j}_{p} \: p \: n_d \right) \: P_j + D^{j+1}_{n} \: n \: n_d \: P_{j+1} \nonumber \\
		& &  + D^{j-1}_p \: p \: n_d \: P_{j-1} \nonumber \\
		\frac{dp}{d\tau} & = & - \sum_{j} D^{j}_{p} \: p \: n_d \: P_j \label{eq:p}
	\end{eqnarray}
	
	\begin{figure}[ht]
		\includegraphics[width=1\columnwidth]{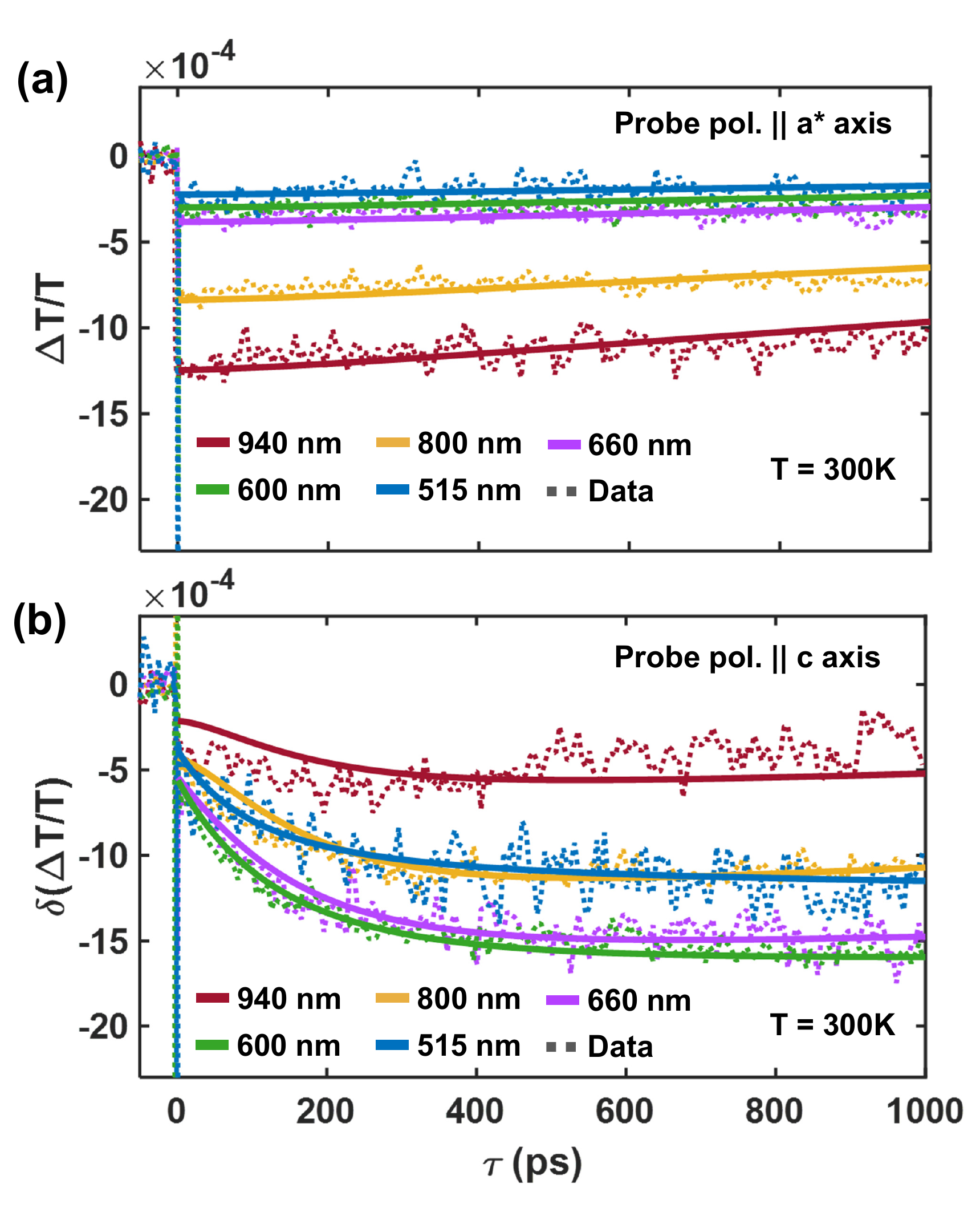}
		\caption{(a) The measured (dashed) and the computed (solid) $\Delta T/T$ transients for probe polarized along the a*-axis, the case in which only free-carrier absorption contributes to $\Delta T/T$, are plotted for different probe wavelengths. (b) The measured (dashed) and the computed (solid) $\delta(\Delta T/T)$ transients, to which only the defect absorption contributes, are plotted for probe polarized along the c-axis for different probe wavelengths.}
		\label{fig:fits}
	\end{figure}
	Here, $n$ (or $p$) is the density of electrons in the conduction band (or of holes in the valence band), $D^{j}_{n}$ (or $D^{j}_{p}$) is the electron (or hole) capture rate constant for the defect in the $j$ charge state. Given that we don't see the experimentally measured absorption peak (centered at 2.2 eV) corresponding to the $0$ charge state decrease with the pump-probe delay, we assume that charge states $-2$ and $-3$ have negligibly small probabilities within the maximum pump-probe delay ($\sim$1 ns) possible in our experiments. Therefore, we adjust the parameters in the equation for $P_{-2}$ such that $P_{-2}$ remains zero in the first $\sim$1 ns. The various processes corresponding to the above rate equations are depicted in Fig. \ref{fig:scheme}. The electron and hole capture rate constants, the defect density $n_{d}$, the initial values $P_{j}(\tau = 0)$ ($j=+1,0,-1$), and the peak values of the defect absorption cross-sections $\sigma_{j}(\omega)$ ($j=+1,0$) are the fitting parameters in the model and their values are chosen to fit the measured $\Delta T/T$ transients for both probe polarizations, for all probe wavelengths, and for all probe delays. The free-carrier absorption cross-section $\sigma_{fc}(\omega)$ is determined as discussed in our earlier work \cite{singh_fc}.   

    Fig.~\ref{fig:fits}(a) shows the measured and the computed $\Delta T/T$ transients for probe polarized along the a*-axis, the case in which only free-carrier absorption contributes to $\Delta T/T$, for different probe wavelengths. Fig.~\ref{fig:fits}(b) shows the measured and the computed $\delta(\Delta T/T)$ transients, to which only the defect absorption contributes. As mentioned earlier, $\delta(\Delta T/T)$ is obtained by subtracting the measured ${\Delta}T/T$ along the a*-axis from that along the c-axis. The parameter values used to fit the data are given in Table II. Fig.~\ref{fig:fits} shows that the model fits the data very well for all probe wavelengths and polarizations, and at all probe delays. From the fits, the defect density was found to be $\approx$ $1.6\times10^{15}$ $cm^{-3}$. The hole capture rate constants are found to be larger than the electron capture rate constants. This is why the maximum magnitude of $\delta(\Delta T/T)$ occurs at $\tau\:>\:0$, long after the pump has passed. The parameter values in Table II are similar in magnitude to the ones determined by Koksal \textit{et al.} for defect states in Sn-doped $\bar{2}01$ $\beta$-Ga$_2$O$_3$ using a much simpler model \cite{koksal_opop}. 
    
	\begin{table}
		\begin{ruledtabular}
			\bgroup
			\def\arraystretch{1.5}
			\begin{tabular}{ccc}
				\textit{Parameter}&\textit{Value}&\textit{Unit}\\
				\hline
				$n_d$ & $\left(1.6\pm1\right)\,\times\,10^{15}$ & $cm^{-3}$ \\
				$D^{+1}_{n}$ & $\left(2.2\pm0.5\right)\,\times\,10^{-9}$ & $cm^{3}/s$ \\
				$D^{0}_n$ & $\left(1.1\pm0.5\right)\,\times\,10^{-10}$ & $cm^{3}/s$ \\
				$D^{0}_{p}$ & $\left(2.8\pm0.5\right)\,\times\,10^{-7}$ & $cm^{3}/s$ \\
				$D^{-1}_p$ & $\left(2.8\pm0.5\right)\,\times\,10^{-7}$ & $cm^{3}/s$ \\
				$\sigma_{+1}|_{peak}$ & $\left(7.9\pm1\right)\,\times\,10^{-17}$ & $cm^2$ \\
				$\sigma_0|_{peak}$ & $\left(5.3\pm1\right)\,\times\,10^{-17}$ & $cm^2$ \\
			\end{tabular}
			\egroup
		\end{ruledtabular}
		\label{tab:params}
		\caption{Extracted values of model parameters}
	\end{table}

\begin{figure}[ht]
		\includegraphics[width=0.85\columnwidth]{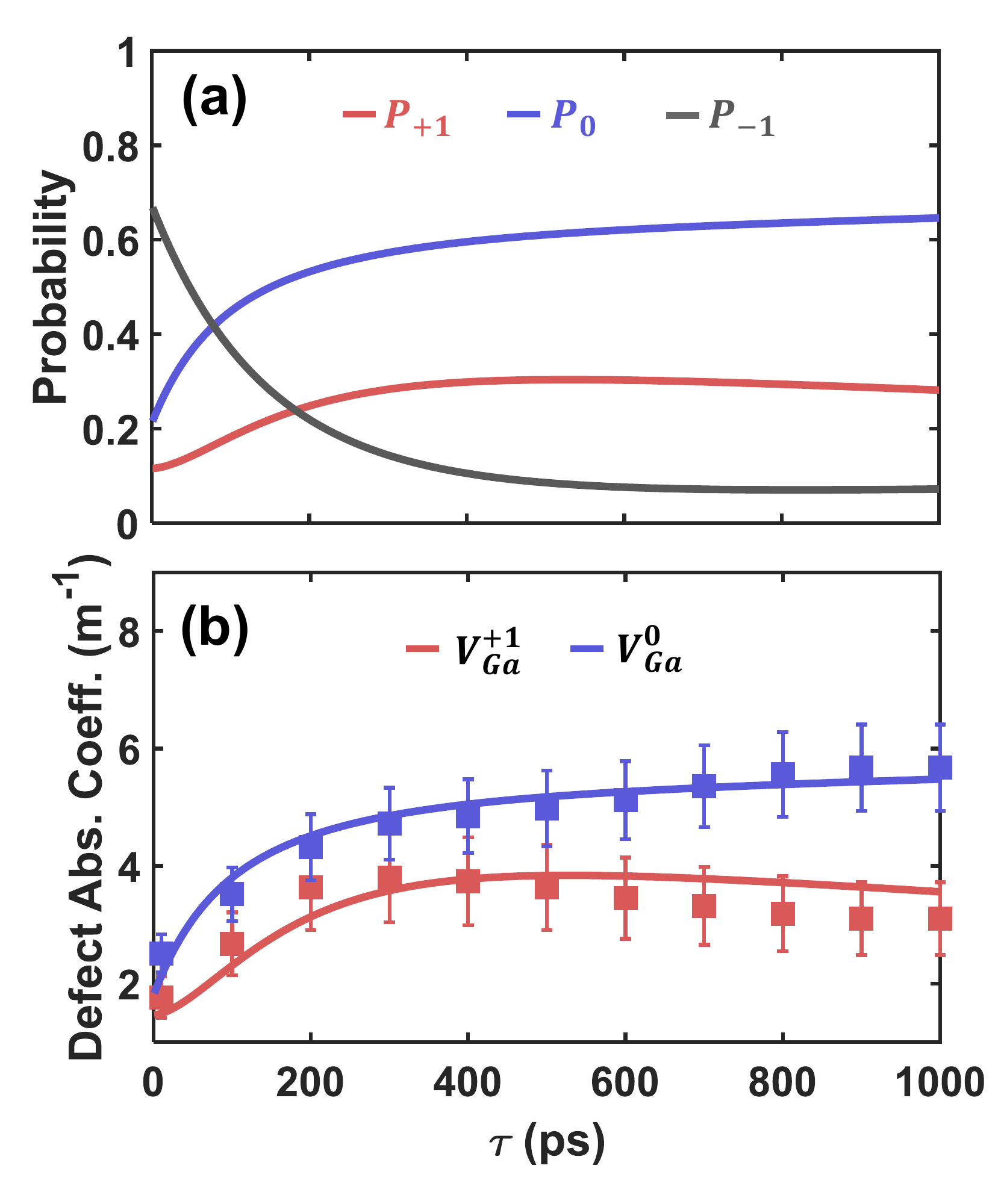}
		\caption{(a) The computed values of the probabilities $P_{+1}$, $P_0$, and $P_{-1}$ for the defect (Gallium vacancy) to be in charge states $+1$, $0$, and $-1$, respectively, are plotted as a function of the probe delay, $\tau$. (b) The computed values of the products $n_{d}\sigma_{1}|_{peak}P_{1}(\tau)$ and $n_{d}\sigma_{0}|_{peak}P_{2}(\tau)$, which are the peak optical absorption coefficients for the defect charge states $+1$ and $0$, respectively, are plotted as a function of the probe delay (solid lines). Also plotted are the values of these absorption coefficients extracted directly from the data plots shown in Fig.~\ref{fig:defect_transmission} (solid squares).}
		\label{fig:model}
	\end{figure}

	Fig.~\ref{fig:model}(a) shows the computed values of $P_{+1}$, $P_0$, and $P_{-1}$ plotted as a function of the probe delay, $\tau$. The values of the products $n_{d}\sigma_{+1}|_{peak}P_{+1}(\tau)$ and $n_{d}\sigma_{0}|_{peak}P_{0}(\tau)$, which are the peak absorption coefficients for charge states $+1$ and $0$, respectively, can also be extracted directly from the data plots shown in Fig.~\ref{fig:defect_transmission}. Fig.~\ref{fig:model}(b) plots these extracted values (solid squares) along with those computed using the rate equations (solid lines). The agreement between the data and the model is again very good.
	
\section{Conclusion}
	In conclusion, we reported experimental results from ultrafast optical-pump supercontinuum-probe spectroscopy of non-equilibrium defect absorption in $\beta$-Ga$_2$O$_3$. Our experimental and theoretical results show that the measured absorption features are due to optical transitions from the valence band to different charge states of Ga(I) vacancies. Good agreement between our first principles calculations and the experimental data, and the ability of our rate equations to model the measured transients for different probe wavelengths and polarizations at all probe delays show that our model captures the underlying physics well. Our results also demonstrate that broadband ultrafast supercontinuum spectroscopy can be a valuable tool to explore defects states and defect dynamics in semiconductors.  
	
	% ---------------
	% acknowledgement
	% ---------------
	\begin{acknowledgments}
		This work made use of the Cornell Center for Materials Research Shared Facilities which are supported through the NSF MRSEC program (DMR-1719875). This work was also supported by AFOSR under Grant No. FA9550-18-1-0529. Computing resources were provided by the Extreme Science and Engineering Discovery Environment (XSEDE), which is supported by NSF grant number ACI-1548562. The authors acknowledge helpful discussions with Dr. Shin Mou (AFRL). 
	\end{acknowledgments}

	% ---------------------------
	% data availability statement
	% ---------------------------
	\section*{Data Availability Statement}
	The data that support the findings of this study are available from the corresponding author upon reasonable request.

	\bibliography{ga2o3_opsp_citations}

\end{document}